\documentclass[runningheads]{llncs}

\usepackage{tikz}
\usepackage{amsmath}
\usepackage{enumitem}
\usepackage{url}
\usepackage{filecontents}
\usepackage{longtable}
\usepackage{graphicx}
\usepackage{amsmath,amssymb}
\def\argmax{\mathop{\rm argmax}}
\usepackage{filecontents}

\usepackage{tikz}
\usepackage{amsmath}

\begin{document}

\title{Change-Point Analysis of Cyberbullying-Related Twitter Discussions During COVID-19}

\def\plainauthor{Sanchari Das, Andrew Kim, Sayar Karmakar}

\titlerunning{Change-Point Analysis of Cyberbullying-Related Twitter Discussions During COVID-19}

\author{Sanchari Das\inst{1,2} \and
Andrew Kim \inst{1} \and
Sayar Karmakar\inst{3}}

\authorrunning{Das et al.}

\institute{Indiana University \and
University of Denver \and
University of Florida \and
}

\maketitle

\begin{abstract}

Due to the outbreak of COVID-19, users are increasingly turning to online services. An increase in social media usage has also been observed, leading to the suspicion that this has also raised cyberbullying. In this initial work, we explore the possibility of an increase in cyberbullying incidents due to the pandemic and high social media usage. To evaluate this trend, we collected $454,046$ cyberbullying-related public tweets posted between January 1$^{st}$, 2020 -- June 7$^{th}$, 2020. We summarize the tweets containing multiple keywords into their daily counts. Our analysis showed the existence of at most one statistically significant changepoint for most of these keywords, which were primarily located around the end of March. Almost all these changepoint time-locations can be attributed to COVID-19, which substantiates our initial hypothesis of an increase in cyberbullying through analysis of discussions over Twitter.
\end{abstract}
\section{Introduction}
\label{sec:intro}

Cyberbullying has become more prevalent, as targeted victimization has moved from in-person to digital platforms, reaching users regardless of geographic constraints~\cite{ybarra2007examining,slonje2008cyberbullying}. Victims of cyberbullying can be targeted through various sources, including mobile phones, video cameras, emails, and web pages~\cite{ybarra2004online}. Targets of cyberbullying-- particularly adolescents-- are more likely to show signs of depression, anxiety, and, in some cases, suicidal behavior~\cite{kowalski2013psychological,schenk2012prevalence,ybarra2004linkages}. Online harassment can carry into adulthood, with bullied victims being more likely to show mental health problems later on~\cite{bannink2014cyber,copeland2013adult}. Such online harassment can negatively impact mental health, with 32\% of victims reporting symptoms of stress and 38\% of victims experiencing emotional distress, even after the online abuse stopped~\cite{finkelhor2000online,ybarra2006examining}. Thus, it is critical to detail cyberbullying and understands the victims' perspectives.

Social media privacy and security have been a concern for many researchers and industry practitioners~\cite{dev2018understanding,dev2019personalized,das2019privacy}. Researchers have often noted that users experience several privacy-focused issues in these social media platforms, which can also lead them to leave such platforms~\cite{noman2019techies}. Privacy policy and recommended changes to the same addressed some of the users' concerns~\cite{das2018modularity,dev2018privacy}, but prior studies have shown that social media usage has increased the extent of cyberbullying~\cite{ybarra2006examining}. On social networking sites and applications, cyberbullying is particularly common, with 66\% of all incidents on these platforms~\cite{pew2014}. Platforms such as Twitter allow people to sometimes interact with strangers (including celebrities)~\cite{das2017celebrities}; however, this also leads others to imitate and forge identities online and trick users~\cite{tsoutsanis2012tackling}.

Furthermore, with the current COVID-19 pandemic, people have increased their social media usage to seek information and stay connected with others while social distancing~\cite{wiederhold2020social}. Social media can be used to support others during crises~\cite{palen2008online}. However, there have also been reports of incivility through such platforms~\cite{kim2020effects}. A sudden rise in social media usage-- combined with children and adolescents regularly using such platforms-- could create a spike in cyberbullying~\cite{digitaltrends2020}. Thus, we specifically wanted to see whether that is the case and answer the following research question: \textit{How do crises, such as a global pandemic (COVID-19), impact cyberbullying trends over social media?}

To understand users' perspectives and the impact of COVID-19 on cyberbullying, we collected $454,046$ of publicly available tweets about cyberbullying to understand user experiences online. As hypothesized, we noticed an increase in cyberbullying incidents and discussions about it during the pandemic. After discussing the impact of cyberbullying and some related works in section~\ref{sec:related}, we provide a detailed methodology, analysis, and findings in section~\ref{sec:method}. We briefly discuss the data collection, pre-processing the data in meaningful categories, and give an overview of the change-point analysis.

\section{Related Work}
\label{sec:related}
Cyberbullying is a major concern for digital communication that can lead to critical consequences. Cyberbullying has increased, given the advent of social media and billions of users being online everyday~\cite{slonje2008cyberbullying}. Additionally, because times of crisis can increase users' online presence and, as a result, cyberbullying, it is important to consider human factors to protect users during such situations-- especially with the current pandemic situation~\cite{depoux2020pandemic}.

\subsection{Cyberbullying}
Mason defined cyberbullying as \lq\lq. An individual or a group willfully using information and communication involving electronic technologies to facilitate deliberate and repeated harassment or threat to another individual or group by sending or posting cruel text and/or graphics using technological means\rq\rq~~\cite{mason2008cyberbullying}. The source of the attack can vary from mobile phones to personal computers to other digital mediums. While studying the various sources of cyberbullying, it is critical to study the behavior and reaction of the attackers and their victims. Nocentini et al. studied the behavior of attackers for different types of cyberbullying, including an imbalance of power, intention, repetition, anonymity, and publicity~\cite{nocentini2010cyberbullying}. 

\subsection{Impacts of Cyberbullying}
Previous works have explored the effects of cyberbullying on targets, especially teenagers; sometimes, such abuse can impact both the cyberaggressors and cybervictims. Bonanno and Hymel found that both victims and perpetrators of cyberbullying were more likely to develop depression and suicidal thoughts than those involved in other types of bullying~\cite{bonanno2013cyber}. Dredge et al. noted the detrimental effects of cyberbullying on targets' social and emotional lives, with the severity of the impact of the harassment depending on different factors, including the anonymity of the perpetrators and bystanders' presence \cite{dredge2014cyberbullying}. Similarly, Wisniewski et al. noted that lower online risk could help in the teens' developmental stages while developing and enhancing crucial interpersonal skills, such as boundary setting, conflict resolution, and empathy~\cite{wisniewski2016dear}. In addition to the mental impact, \v{S}l{\'e}glov{\'a} and Cerna found that cyberbullying led to behavioral changes, with victims displaying more cautious browsing habits and avoidance strategies~\cite{vsleglova2011cyberbullying}. McHugh et al. noted the negative emotions victims of cyberbullying experience, though they also found that the impact may be more short-term than previously thought, emphasizing the importance of resilience~\cite{mchugh2017most}.

\subsection{Social Media Bullying}
\label{sec:socialbullying}
Cyberbullying can occur across a range of different online platforms, including social networks, chat rooms, and mobile messaging applications, regardless of geographic proximity; such bullying can last as little as a week or go on for much longer \cite{smith2006investigation}. Because social networking platforms are often used as a means of self-comparison, they are a prime source of self-esteem issues~\cite{vogel2014social}. Several high-profile incidents of cyberbullying have taken place over major social media platforms. In May 2020, a Japanese reality TV star took her own life after being subject to abuse on social media~\cite{bbc2020}. Similar incidents across the world have led lawmakers to pass legislation that would make cyberbullying criminal~\cite{hinduja2015cyberbullying}.

As a mitigating measure, some prior work has focused on improving social media policies to prevent perpetrators from abusing their victims. Milosevic examined the responsibilities of social media companies' in addressing cyberbullying among children~\cite{milosevic2016social}. They mention concerns on the transparency and accountability of these platforms in addressing and mitigating such issues. 

\subsection{Cyberbullying Trend Analysis}
Studies that analyze trends in cyberbullying help understand how events can impact digital users. Schneider et al. conducted four surveys across 17 high schools and found that the overall rate of cyberbullying increased from 2006 to 2012~\cite{kessel2015trends}. Through survey-based analysis, Snell and Englander found that females are more likely to be involved in cyberbullying as both victims and as perpetrators, indicating the importance of gender as a factor in mitigating online bullying~\cite{snell2010cyberbullying}. Mangaonkar et al. used a distributed design for analyzing tweets and detecting cyberbullying in real-time~\cite{mangaonkar2015collaborative}.

Twitter allows users to express themselves in 280 character \lq tweets;\rq~ prior studies have analyzed these messages for cyberbullying~\cite{alim2015analysis,nurrahmi2018indonesian}. Cortis and Handschuh analyzed bullying tweets in the context of two trending events (the Ebola outbreak and the shooting of Michael Brown in Ferguson, Missouri). They identified commonly used hashtags and named entities in bullying tweets~\cite{cortis2015analysis}. Whether or not such crises increased, bullying tweets were not studied. Due to an increase in individuals' online digital presence, assumptions have been made that the pandemic situation from COVID-19 can increase cyberbullying incidents. Thus, our goal is to understand the trend and find evidence to support or contradict this hypothesis.

\section{Methodology and Findings}
\label{sec:method}
Twitter is a social networking site where users can post real-time messages. With over 300 million active daily users, it is an ideal data source~\cite{statista2019}. To assess the impact of COVID-19 on cyberbullying, we collected $454,046$ public tweets on Twitter, all of which mentioned cyberbullying. We outline our process for collecting and analyzing the relevant tweets below.

\subsection{Data Collection}
We scraped Twitter for user-posted, publicly available tweets related to the topics of cyberbullying, social media bullying, online harassment, etc. More specifically, we used the following key terms when conducting our search:~\textit{ Internet bullying, Internet bully, Internet bullies, online abuse, online harassment, online shaming, online stalking, cyberbullying, social media bullying, stop cyberbullying, cyber bully, cyber bullies, FB bullying, FB cyberbullying, FB harassment, FB victim, Facebook bullying, Facebook cyberbullying, Facebook victim, Facebook harassment, Twitter bullying, Twitter cyberbullying, Twitter harassment, Twitter victim, Insta bullying, Insta cyberbullying, Insta harassment, Insta victim}.

The data was collected using the Get Old Tweets API~\cite{henrique2016}, which allowed us to access tweets older than one week. This API was used in the web crawler, written in Python, and the data was stored with MongoDB. The data collection spanned from January 1$^{st}$, 2020--June 7$^{th}$, 2020. This timeline was mainly selected to note the impact of COVID-19 on online users and determine whether the crisis led to an increase in online abuse. We specifically only collected direct tweets and removed any retweets or duplicate tweets.

\subsection{Analysis}
\begin{figure*}[ht]
    \centering
    \includegraphics[width=\textwidth]{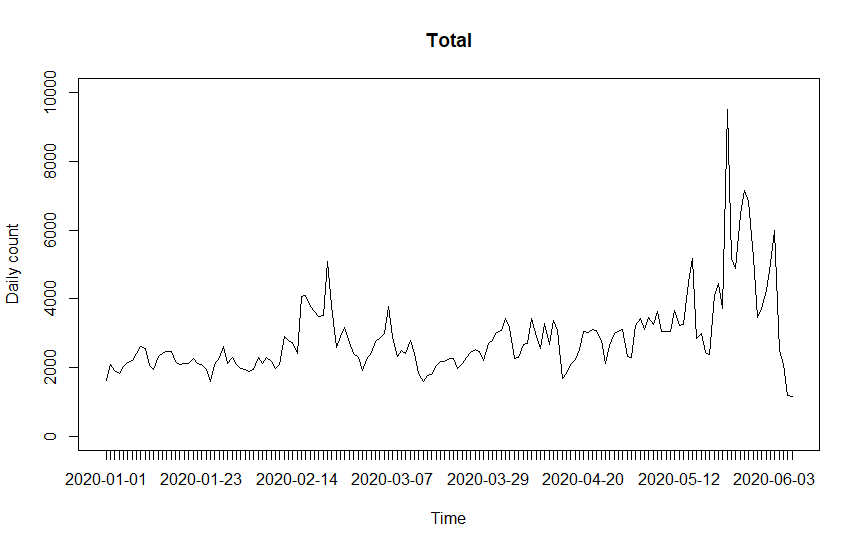}
          \caption{Daily count of total tweets related to cyberbullying}
    \label{fig:tc}
\end{figure*}

After completing the data collection, we performed a trend analysis to evaluate the impact of the crisis. Using the timestamp of the post, we obtained the daily count of the tweets, which including at least one of these keywords. Figure \ref{fig:tc} shows the daily count for the $159$ days from 01$^{st}$ January, 2020 to 07$^{th}$ June, 2020.

\begin{figure*}[ht]
    \centering
    \includegraphics[width=\textwidth]{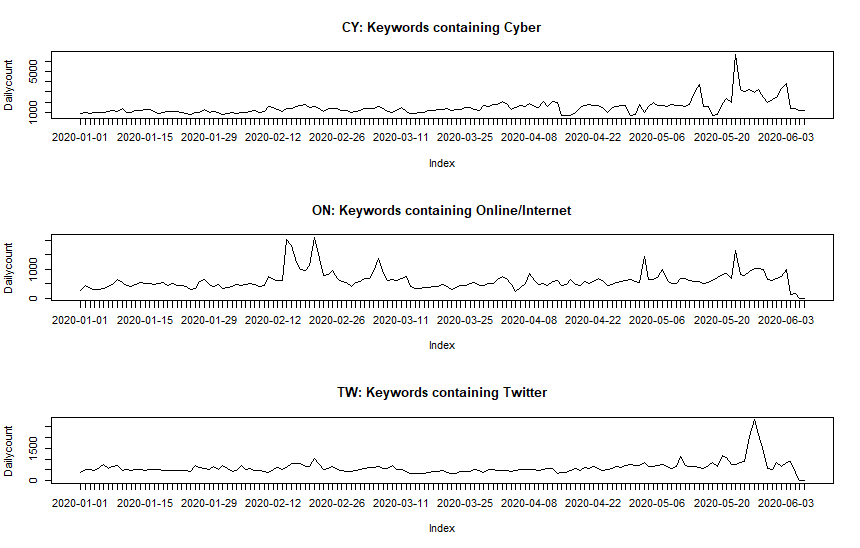}
          \caption{Daily count of total tweets for the three sub-classes (CY, ON, and TW)}
    \label{fig:sub-class}
\end{figure*}

There were 28 different keywords (mentioned above). Some of the keywords had fewer tweets with a negligible impact on the analysis. Thus, we broadly divide them into three sub-classes: keywords containing "cyber" (CY) for the keywords -- cyberbullying, cyber bully, cyber bullies, stop cyberbullying, FB cyberbullying, Facebook cyberbullying, and Insta cyberbullying; "online/internet" (ON) for the keywords -- internet bullying, Internet bully, Internet bullies, online abuse, online harassment, online shaming, online stalking, and "twitter" (TW) for keywords -- Twitter bullying, Twitter cyberbullying, Twitter harassment, and Twitter victim. The total daily counts for these sub-classes are shown in Figure~\ref{fig:sub-class}. We also tabulate them later in the changepoint analysis in Section~\ref{sec:changepoint} (Table \ref{tab:sub-class}).

\begin{figure*}[ht]
    \centering
    \includegraphics[width=\textwidth]{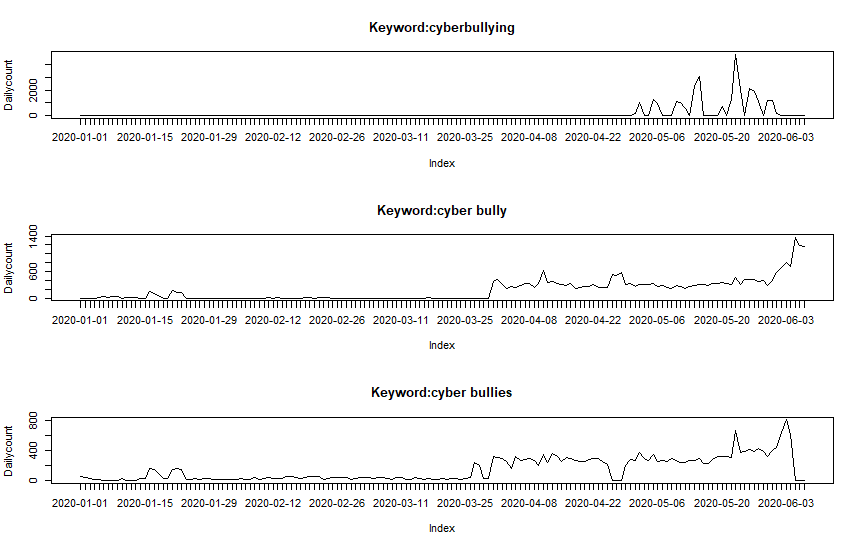}
          \caption{Impact of the Cyberbullying Incidents dependent on Three Major Keywords (cyberbullying, cyber bully, and cyberbullies)}
    \label{fig:interesting}
\end{figure*}

One can see a pattern prevalent to all the counts and the sub-classes that we present here. Overall, except for the sub-class \lq ON\rq~, there does not seem to be a considerable change in mean except it went slightly upwards since mid-March and in all categories, including the total. We notice a more considerable spike in the cyberbullying related tweets in the second half of May. The sudden rise in the frequency of tweets in the second half of May can be due to the suicide of the Japanese TV star~\cite{michel2020}. Moreover, for the class \lq ON\rq~, one can see a spike in the second half of February, and the overall mean also had an upward trend. This may or may not be due to the pandemic. Since these are cumulative graphs of the prevalence of such words, we wanted to introspect in each of the 18 keywords in the subclasses (CY=seven, ON=seven, TW=four keywords). Out of them, we observed three keywords: \lq\lq cyberbullying, cyber bully, cyber bullies\rq\rq~ having a significant impact which we summarize in Figure~\ref{fig:interesting}. We provide some mathematical details in the next subsection about how to formally test changepoint.

\subsection{Changepoint Analysis}
\label{sec:changepoint}
Assume we observe $X_1, \ldots, X_T$ over $T$ time-points, and we are suspecting at most ONE changepoint (AMOC) location $\tau$ in mean i.e. 
$$ E(X_i)= \mu\text{ if }i \leq \tau, \text{ and } E(X_i)= \mu+\delta\text{ if }i >\tau \text{ where } \delta \neq 0.$$  

\noindent In layman's term this mean that the realized counts as random variables have a different mean before and after the change-point location $\tau$. Statistically speaking this difference needs to be significant to be able to be detected from the observed data. We adopt a CUSUM technique for estimating the changepoint location. We define it as:

\begin{eqnarray}\label{eq:cpdefine}
\hat{\tau}= \argmax_{1 \leq s \leq T} \frac{s(n-s)}{n} \left(\frac{1}{s}\sum_{i=1}^s{X_i}- \frac{1}{n-s}\sum_{i=s+1}^T{X_i} \right)^2
\end{eqnarray}

Intuitively, the above equation is the location that maximizes the difference of normalized cumulative sum before and after this point. There have been previous literature on offline change-point detection~\cite{page,hinkley,petit}. Here, for simplicity, we assume independence over time-horizon. In statistics literature, consistency results typically assume the observations to be Gaussian; however, since this is a count data, it can be questionable. In light of the weak law of large numbers, however, one can assume normality as the counts are large. To detect the changepoints, we employ the \textit{changepoint} package in R and observe the following changepoints in the three sub-classes and the three significant individual keywords. These are tabulated in Tables~\ref{tab:sub-class} and~\ref{tab:keyword}. All these changepoints were significant at the type-1 error level $\alpha=0.05$.

\begin{table}[ht]
\centering
\begin{tabular}{|r|r|r|r|r|}
  \hline
 Name & Contains & \#keywords & \#tweets & Changepoint\\ 
  \hline
   CY & "Cyber" & 7 & 235,542 & 29th March\\ 
   ON & "Online" & 7 & 96,629 & 11th Feb\\ 
   TW & "Twitter" & 4 & 96,147 & 27th April\\ 
  \hline
\end{tabular}
\vspace{5mm}
\caption{Changepoint analysis of 3 sub-classes}
\label{tab:sub-class}
\end{table}

\begin{figure*}[ht]
    \centering
    \includegraphics[width=\textwidth]{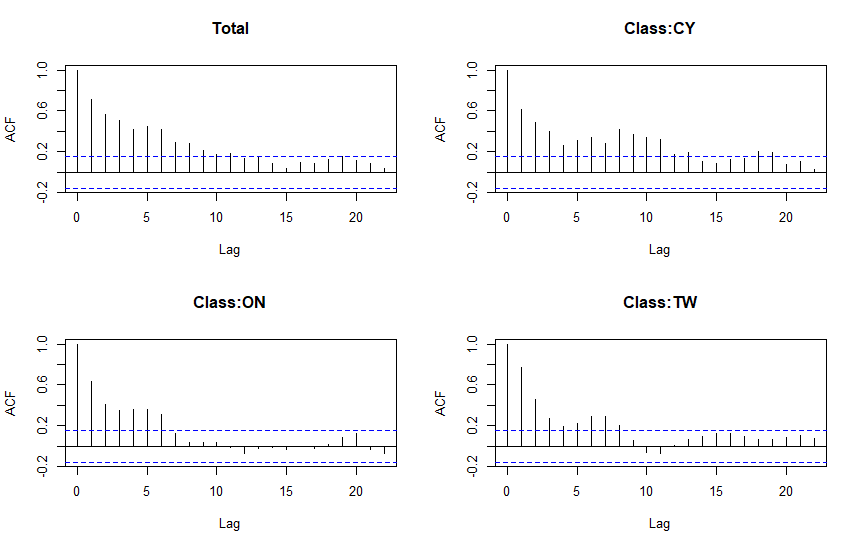}
          \caption{Auto correlations for the total count and  sub-classes}
    \label{fig:acfcb}
\end{figure*}

\begin{table}[ht]
\centering
\begin{tabular}{|r|r|r|}
  \hline
  Keyword & \#tweets & Changepoint\\ 
  \hline
   "cyberbullying" & 29,477 & 2nd May \\
   "cyber bully"  & 27,806 & 31st March \\
   "cyberbullies" & 24,287 & 31st March \\
   \hline
\end{tabular}
\vspace{5mm}
\caption{Changepoint analysis of 3 specific keywords}
\label{tab:keyword}
\end{table}
We note that, for all the series and sub-classes we observe a changepoint. Except for the subclass `ON,' all of the changepoints can be possibly be attributed to COVID. However, the total count did not show any changepoint, and we think this can be due to multiple reasons. First, we are adding a lot of keywords. Thus, the effects might get confounded. A more important reason could be the simplicity of the assumption of independence. We show in Figure~\ref{fig:acfcb} that the total count of the individual keywords and all three sub-classes show significant correlation over time. Once the dependence is taken care of, it is possible that even the total count data will show changepoints somewhere around the end of March. We wish to explore this as a future work. 

\section{Conclusion and Future Works}
\label{sec:conclusion}
In this work in progress, we wish to explore a comprehensive time-trend analysis of the impact of COVID-19 on cyberbullying as suspected by many experts. We found that certain class of keywords show a change in cyberbullying related tweets from the end of February or March when the pandemic fear primarily started. As a future extension of this work, we would like to comprehensively address this using a change-point analysis for a time-series of count data. One can also implement possible changes in variance since we observed some fluctuations in the tweet counts. An interesting finding from this initial analysis is that the change points for different sub-class and tweets are not necessarily close. This can lead us to employ methods from prior work Karmakar et al. ~\cite{karmakarcpsync} to statistically validate the hypothesis of synchronization of changepoints, as the authors therein allowed for non-linear non-Gaussian time-series.

We are also working on an alternative formulation of the same problem using a Bayesian time-varying paradigm~\cite{karmakarevaluating2020}. We assume the parameters of the models do not change abruptly if there is a change-point but instead shows a more gradual change. We wish to explore time-varying models in a frequentist sense as done in~\cite{sayar2020,karmakar2018} or Bayesian methods from a relatively recent work by Roy and Karmakar~\cite{sk2020} in the regime of count autoregressive series. This would allow us to incorporate dependence in the analysis and give a clearer picture of how the mean or the dependence coefficient changed over time (and thus if COVID-19 had a telling effect on the increase). A time-series formulation often asks for prediction of the future, and such a work has not yet been done in the field of cyberbullying trend analysis. Instead of a single $k$-step ahead forecast, we would like to predict the trend for an entire month or two. We wish to explore statistical methods developed by Zhouwu et al. and Chudy et al.~\cite{zhouwu2010,chudy2020} to this non-Gaussian count time series and build statistically valid prediction intervals. This can help create a mitigating strategy in case we can predict a rise of cyberbullying for the next one or two months. 

\section{Acknowledgments}
We would like to thank Umang Mehta for his help with the data collection. We would also like to acknowledge the research institutions and labs of the researchers involved with this project- Secure and Privacy Research in New-Age Technology (SPRINT) Lab, University of Denver; and Human and Technical Security (HATS) Lab, Indiana University, and the University of Florida. Any opinions, findings, and conclusions or recommendations expressed in this material are solely those of the author(s). 

\bibliographystyle{plain}
\bibliography{Bully.bib}

\end{document}